# Space Elevator Propulsion by Mechanical Waves

Mark A. Wessels, Ph.D.
Collin County Community College
Frisco, Texas
February 21, 2018

**Abstract**
The current preferred envisioned method for transmitting power to a space elevator climber is a laser / photovoltaic (PV) system. In this, a ground-based laser beam would transmit megawatts of optical power through the atmosphere to an arrangement of PV panels mounted on the ascending climber. Although this technique has been successfully demonstrated in small models, this method will likely suffer from serious shortcomings in a realistic full-scale system, including *poor conversion efficiency*, *obscuration by clouds*, and *mechanical fragility* of the panels; worse, the PV method provides *no means* of regenerative energy recovery. Furthermore, the laser would need to operate continuously at multi-megawatt levels for as long as 14 days (the time for the climber to reach geosynchronous altitudes). No such laser has ever been demonstrated.

This paper presents a radical alternative method for propelling a space elevator car: by using the cable to transmit power in the form of *transverse mechanical waves* propagated on the cable. A ground-based mechanical driving oscillator would excite the waves. Traveling upward at hypersonic speeds, they encounter the climber. This mechanical power is then extracted by an engine in the climber to propel the climber upward. The oscillator may be manifested by a pair of opposing pistons contacting the cable on opposite sides, or by an electromagnetic (EM) driver powered by electromagnets located on opposite sides of the cable, near the anchor point. Most importantly, *existing engines* can easily provide the required amount of power to send a 10 metric-ton car from the ground to geosynchronous altitudes.

**Introduction**
The past 20 years have seen a resurgence of interest in the Space Elevator (SE) concept. One major engineering problem to solve is how to provide power to the climbers. Several possibilities include: 1) transmitting *electric currents* in the cable; 2) having an *internal power source* such as a miniaturized nuclear reactor; and 3) using an *optical transmission / conversion system*, whereby a powerful, highly-collimated beam of light (preferably generated by a laser) would be focused onto PV panels attached to the climber. These PV panels would convert the radiant power directly into electricity for driving motors. Any successful method must be capable of transmitting several megawatts of power at reasonable system efficiencies.

In order to transmit *electric currents* at high efficiency, the cable itself must be an extremely good conductor of electricity. The currents would need to be conducted over many thousands of miles / kilometers (geosynchronous altitude is 22,300 miles or 35,900 km high). *Individual* carbon nanotubes have been demonstrated to possess *both* the required mechanical properties necessary for the Elevator, and extremely high electrical conductivity. If near-flawless nanotubes can ever be produced at the macroscopic scale, this method could work. However, the actual material used in a real Elevator cable may not conduct electricity well.

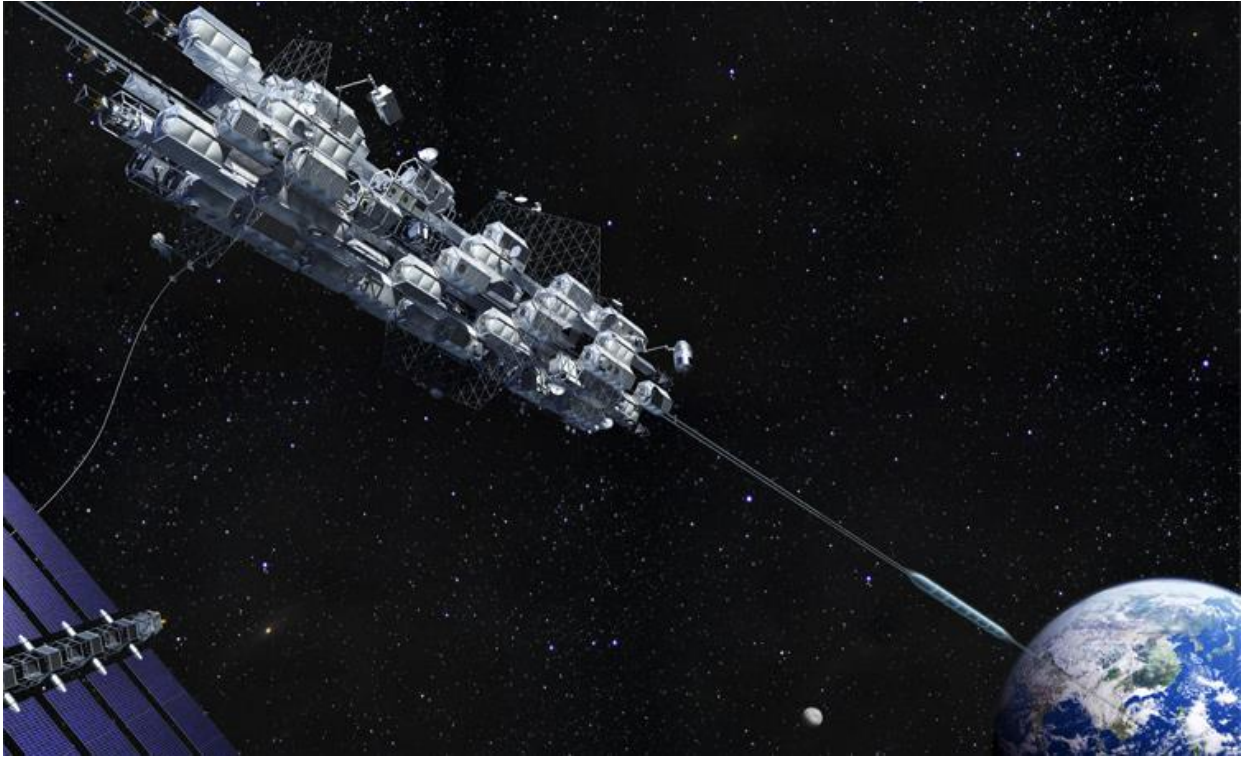

**Figure 1**: **Artistic Rendering of a Space Elevator Concept**. A cable system, some 60,000 miles long and attached to the Earth, would extend into space. Special cars - climbers - would travel along the cable. The Elevator could be built if we can produce large amounts of a material that is both strong and light enough to satisfy the Elevator physics. A much less demanding *lunar* version of the Elevator is also possible, and the materials needed to construct it already exist. (Credit: Obayashi Corporation)

The *internal power source* would not require an electrically conductive cable. Batteries would be out of the question, due to their extremely low energy densities. A miniaturized nuclear reactor, powering a thermo-electric generator - having no moving parts - could provide a workable solution. The conversion efficiency of thermo-electric generators is low, but the reactor would provide so much thermal power that this would not be a problem. The overall weight of the system could still be problematic. This method is unlikely to be compatible with a regenerative energy recovery system, reducing the overall efficiency of the climber system.

The current favorite is the *PV system*. Recent SE prototype competitions have successfully demonstrated PV systems on a small scale. These shine a source of bright light (*not* a laser) upon a PV panel that is attached to the model climber. In reality though, the light-beam method comes with serious shortcomings, including *poor conversion efficiencies* of both the light source and the PV cells; *cloudy weather*, blocking or scattering the beam; *atmospheric turbulence*, which would affect focusing of the beam; *thermal blooming*, which also disrupts focusing; and the *mechanical fragility of the panels* (which could break-off if the climber were to suddenly jam). This method also would not be compatible with an energy recovery system.

A radical alternative to all of these proposed methods is to send the power along the cable in the form of *transverse mechanical waves*, excited and propagated on the cable itself. Waves transmit *energy* across distances, without having to move *mass* across the same. This fact makes wave power ideal for space elevator propulsion.

## Power Requirements

The weight of the climber is its mass *m* multiplied by the acceleration of gravity, *g*, or *mg*. The minimal theoretical *lifting power* P is the *weight* (a force) multiplied by the *vertical velocity* of travel, v. Therefore, P = mgv. A reasonable value for the mass of a climber is 10 metric tons (10,000 kg). A value of 33 meters /sec (74 miles per hour) for velocity seems practical. Thus P = (10,000 kg)(9.8 m/sec$^2$)(33 m/sec) = **3.23 million Watts**, or 4,330 horsepower (hp).  (This paper will not consider system efficiency as a factor, only minimum theoretical values.)

This figure is applicable only for low altitudes, near the surface of the Earth. As the altitude increases, the value of g decreases. This is due to 1) the *reduction in the force of gravity* with increasing altitude, and 2) the *outward centrifugal force that builds* with altitude. Thus, with increasing altitude, the power requirement actually *decreases*.

## Derivation of Wave Power

The following is a derivation of the power transmitted by a transverse mechanical wave traveling in one direction. Let

$$y(x, t) = A\sin(kx - \omega t - \varphi), \text{ (Eq. 1)}$$

where:

y(x, t) is the transverse displacement of the wave as a function of position x and time t,
A is the maximum amplitude,
k is the *wavenumber*, defined as $2\pi / \lambda$, where $\lambda$ is the wavelength,
$\omega$ is the angular frequency ($2\pi f$), and
$\phi$ is an arbitrary phase.

The cable is under a *tension* force, denoted by $\tau$. The *instantaneous transmitted power P* is given by the *transverse force x transverse velocity*, or

$$P = \tau \left|\frac{dy}{dx}\right| \left|\frac{dy}{dt}\right| \quad \text{(Eq. 2),}$$

where |dy/dt| is the magnitude of the transverse velocity, and the product $\tau*|dy/dx|$ gives (for small amplitudes) the transverse force.  Substituting Eq. 1 into Eq. 2 and evaluating the derivatives yields:

$$P = \omega k A^2 \tau \cos^2(kx - \omega t - \varphi). \quad \text{(Eq. 3)}$$

Averaged over one cycle, $\langle\cos^2(u)\rangle = \frac{1}{2}$. Therefore, the *average transmitted power* is given by

$$P_{avg} = \frac{1}{2}\omega k A^2 \tau. \qquad \text{(Eq. 4)}$$

In addition, we have the wave relation k = ω/v, where *v* is the wave velocity. Substituting this relation for k yields

$$P_{avg} = \frac{1}{2}\tau \frac{\omega^2 A^2}{v} \qquad \text{(Eq. 5)}$$

Values for τ, ω and A must be found to yield the required amount of power needed for propulsion.

## Wave Velocity
At the anchor point, the cable tension will have a value near 200,000 Newtons (corresponding to the weight of some 20,000 kg or 44,000 lbs). Near the surface of the Earth, the tension will not deviate much from this value. At greater altitudes, the tension increases dramatically. The speed of the wave can be made both *independent* of the tension and *constant* all along the cable. This can be shown as follows.

The velocity of a small-amplitude traveling wave is given by

$$v = \sqrt{\tau/\mu}, \qquad \text{(Eq. 6)}$$

where μ is the *mass per unit length* of the cable. Let σ be the *cross-sectional area* of the cable at any given position. Dividing both τ and μ by σ yields

$$v = \sqrt{\frac{\tau/\sigma}{\mu/\sigma}}. \qquad \text{(Eq. 7)}$$

The quantity in the numerator of the square root, τ/σ, is the *stress* (force per area). Most SE designs keep the stress *constant* along the cable. The quantity in the denominator, μ/σ, has units of *mass per volume*. This quantity is the *structural density* of the cable material. Most generally, this quantity should also be constant over the entire cable. Therefore, the wave velocity will be constant over the entire cable.

## Design Calculations
Current Elevator designs set the average value of μ at just 15 grams per meter. Thus, the total mass of the cable would be 1,500,000 kg (1500 metric tons) for a cable 100,000 kilometers long. The wave velocity is found to be:

$$v = \sqrt{\frac{200{,}000\ N}{0.015\ kg\ per\ meter}} = \mathbf{3650}\ \text{m/sec. Thus, the wave propagates at \textit{hypersonic} speeds.}$$

A value of **60 centimeters** is chosen for A, the wave amplitude. By solving Eq. 5 for ω, the frequency of oscillation is found to be

$$\omega = \sqrt{\frac{2(3.23 \times 10^6 Watts)(3650\ m/sec)}{(200{,}000\ Newtons)(0.6\ m)^2}} = 572.3 \text{ radians / sec, or } \mathbf{f = \omega / (2\pi) = 92 \text{ cycles / sec}}.$$

This frequency is equivalent to **5520** revolutions per minute (RPMs), within the mechanical abilities of existing engines. The period T is then 1 / 92 second. The wavelength is found to be λ = vT = **39.67 meters**. The *maximum transverse force* (MTF) is seen to be τ*|dy/dx| = τkA = (200,000 N)(2π rad / 39.67 m)(0.6 m) = **19, 000 N**, the weight of 1,939 kg.

## Driving Engine

A mechanical oscillator applies transverse forces to the faces of the cable. The 'cable' is actually a *ribbon*, having two faces. The oscillator includes a support structure, a motor, and a driver. Forces are applied to the ribbon in an alternating cycle. During the first half of the cycle, force is applied to one face of the cable. During the second half, the force on the front face is diminished while force on the back side is applied. This cycle is then repeated. The driver includes two rotating discs (flywheels). Each disc drives a linear piston against the elevator cable. The pistons are set *180 degrees out-of-phase* so that as one piston reaches its maximum advancement, the other reaches its maximum retreat. See Figure 2.

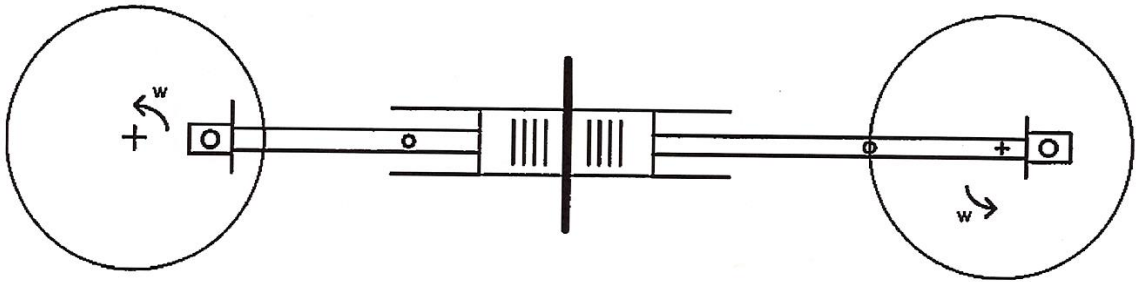

**Figure 2: Opposing Pistons of the Driving Engine**. The circles on either side represent flywheels, which are connected to powerful motors (not shown) capable of generating several thousand horsepower each. Shafts connect the flywheels to pistons, which alternately push against the face of the ribbon. The pistons are set 180 degrees out-of-phase, so that as one advances, the other retreats. This is necessary to produce sinusoidal waves on the ribbon.

To achieve the greatest driving efficiency, the driver is positioned *¼ of a wavelength* above the anchor point. At this point, a *resonance* occurs between the driver and the cable, allowing the maximum transference of mechanical power. See Figure 3.

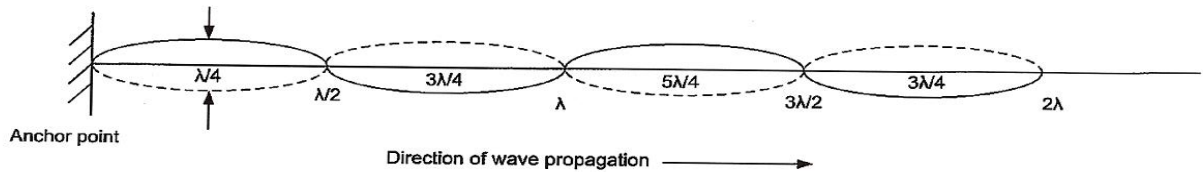

**Figure 3: Propagation of Waves upon the Cable**. The driving pistons push on the ribbon at a point that is ¼ λ above the anchor point. At this position, a *resonance* occurs, a place where the most efficient transference of power occurs.

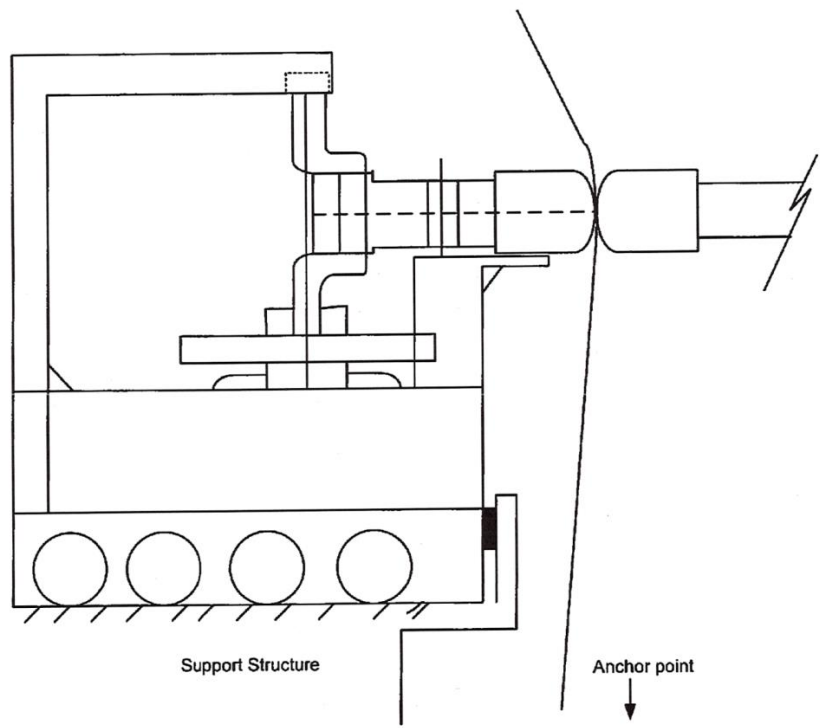

**Figure 4: Side View of Mechanical Driver**. Two opposing drivers are placed across from each other, making contact with the Elevator ribbon at the resonance point, ¼ λ above the anchor point. Pistons are driven by motors that can supply at least 2165 hp each. Each piston is connected to a crankshaft that is displaced from its axis by an amount equal to the amplitude of the generated waves.

# Electromagnetic (EM) Driver

An alternative way of transmitting power onto the cable is with an *electromagnetic driver*. Powerful electromagnets are positioned on either side of the ribbon. Attached to the ribbon at the resonance point is a ferromagnetic material. This material is both attracted and repelled by the magnetic forces generated by each pole face, and would allow flexion of the ribbon at this point. Advancements in material technology might allow *magnetic dopants* to be directly incorporated into the ribbon material itself, thereby avoiding the need to physically attach a large number of individual magnets to the ribbon. See Figure 5.

Precise control of both the frequency and amplitude of the driving forces is possible with an EM-driver. It would have no moving parts. The major concern with an EM driver is *heating by hysteresis*, which could waste large amounts of power. To avoid this problem, the magnet material must be *magnetically soft*. Still, some inefficiency will exist, likely requiring EM drivers to be cooled.

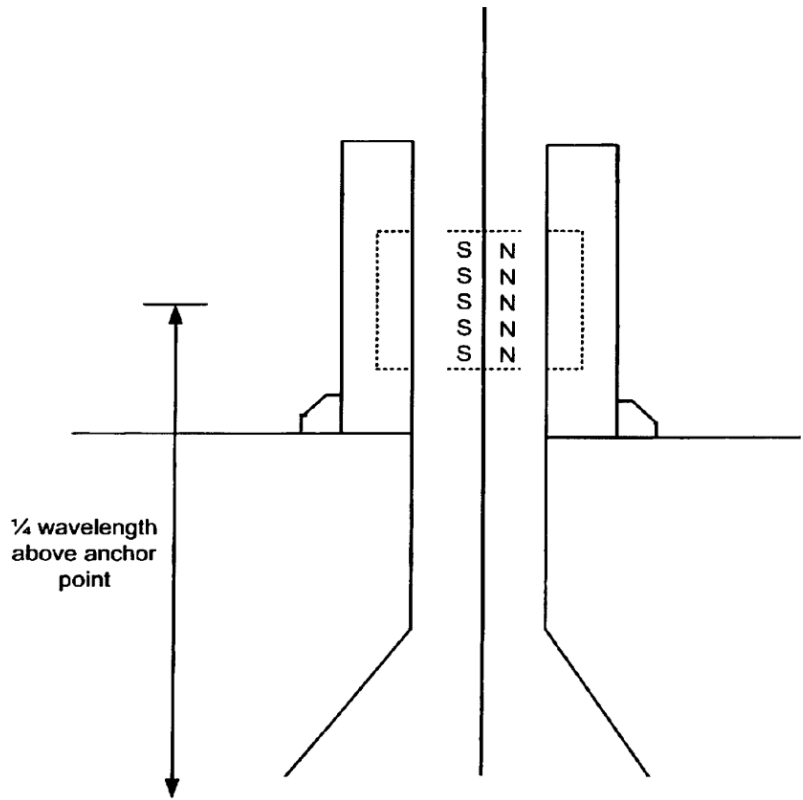

**Figure 5: EM Driver**. Two powerful electromagnets are placed on opposing sides of the ribbon. Powerful alternating electric currents would surge through each pole, producing strong magnetic fields. An array of small permanent magnets could be attached to the ribbon faces, with their north poles (N) on one side, and south poles (S) on the other. The field produced by the electromagnets would alternately push and pull against the ribbon, thereby driving oscillations. Magnetic materials could even be incorporated into the ribbon itself, achieving the same result.

## Climber Transmission / Engine

In order to transfer the thousands of horsepower of mechanical power to the climber, the equivalent of an automobile transmission is needed. This machine must first *extract* mechanical power from the ribbon and then *convert* it to a form that is able to pull the climber upward. Ideally, this transmission would also run in reverse. When the climber descends, its engine would convert its gravitational potential energy into waves that would travel downward to the base. The driving engine could then act as a generator, producing power that could be stored or returned to the grid. An efficient system could recover as much as 90% of the energy initially needed to raise the climber. (The famous science fiction writer Arthur C. Clarke once estimated that a well-designed space elevator system using regenerative methods could lower payload costs to as little as $100 per pound.)

The transmission should work in a way that prevents the formation of *wave reflections*. Reflected waves may disrupt the power transmission process and possibly damage the driver. The physics of wave reflection is completely analogous to that of reflected electromagnetic power in a transmission line (the "standing wave ratio" (SWR) of radio parlance). In this way, the climber's transmission may be considered a *mechanical impedance matcher*.

The most important design consideration for the transmission is the wave *frequency*. To absorb the power in the most efficient manner, the transmission must be "tuned" to the precise frequency of the incoming waves.

Figure 6 illustrates a side view of one possible climber transmission system. The transmission includes a plurality of *extractors* - linearly-aligned pistons, each coupled to its own mass-spring system. Each extractor drives its own electric generator. An extractor includes a piston which is in direct contact with ribbon, and a permanent magnet placed within a coil. The generated electric currents drive an electric motor (or a plurality of motors) attached to the climber, thereby drawing the climber up the cable. The power extraction should be *evenly distributed* across all extractors. The *number* of extractors in the climber is an important design parameter. The physics of how the wave interacts with the extractors may be discussed in future papers.

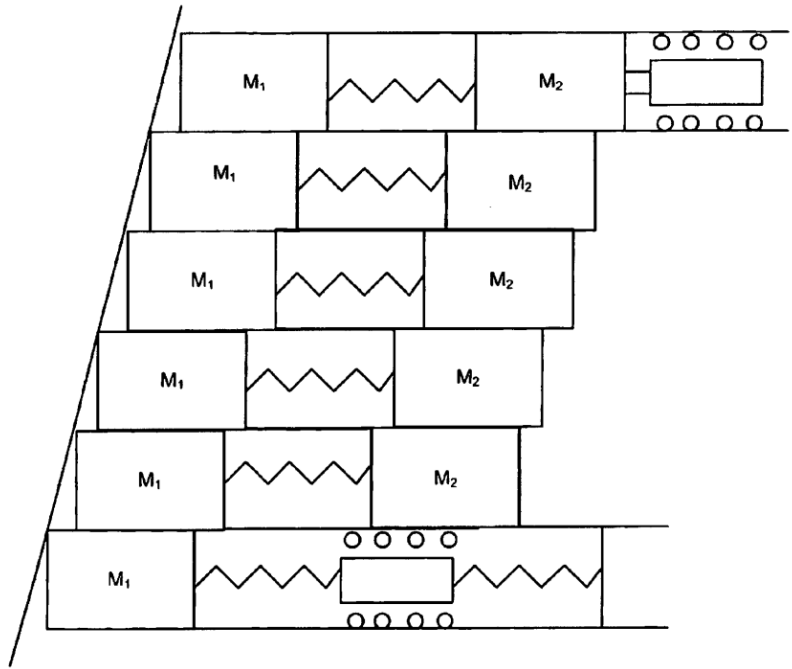

**Figure 6: Climber Transmission Concept**. An arrangement of *extractors* - piston-mass-spring oscillators - extracts the wave power on the ribbon. Each is tuned to the precise frequency of the incoming waves. The slanted line at left represents the ribbon. $M_1$ represents a piston, while the triangle-wave represents a spring. $M_2$ is a second mass, connected to a permanent magnet or, as seen in the bottom-most level, *is* the magnet. The small circles represent a coil of wire. As the magnet oscillates in and out of the coil, electric currents are generated by electromagnetic induction. These currents then drive motors (not shown) to pull the climber upwards. Such a system could also be run in reverse, to recover energy. An identical arrangement is positioned on the opposite side of the ribbon, so that power is extracted on *both* the positive and negative-going phases of a wave cycle.

## Advantages of Mechanical Wave Propulsion
In summary, the wave propulsion method should have the following advantages over other proposed methods of propelling Space Elevator climbers:

1. Driving engines capable of supplying the required amount of power *already exist*. No new technology would be required to build these engines.
2. Climber transmission requires no new technology.
3. System does not require the Space Elevator cable to conduct electricity.
4. System would be largely immune to weather conditions.
5. Compatible with both land and ship-based anchor designs.
6. Compatible with regenerative energy recovery.